\documentclass[12pt,preprint]{aastex}

\newcommand\UCHII{UCH}
\newcommand\HII{H\,{\sc ii}}

\newcommand\kms{km~s$^{-1}$}

\newcommand\cmthree{cm$^{-3}~$}

\newcommand\etal{et al.~}
\newcommand\be{\begin{equation}}
\newcommand\ee{\end{equation}}
\newcommand\bea{\begin{eqnarray}}
\newcommand\eea{\end{eqnarray}}

\newcommand\ddeg{$^{o}$}

\catcode`\@=11
\newcommand{\gsim}{${\mathrel{\mathpalette\@versim>}}$}
\newcommand{\lsim}{${\mathrel{\mathpalette\@versim<}}$}
\newcommand{\@versim}[2]{\lower 2.9truept \vbox{\baselineskip 0pt \lineskip
    0.5truept \ialign{$\m@th#1\hfil##\hfil$\crcr#2\crcr\sim\crcr}}}
\catcode`\@=12

\slugcomment{Submitted to ApJ}

\shorttitle{Carbon recombination line emission in W49 North}
\shortauthors{Roshi, De Pree, Goss \& Anantharamaiah}

\begin{document}

\title{VLA Observations of Carbon 91$\alpha$ Recombination Line Emission in W49 North} 

\author{D. Anish Roshi}
\affil{Raman Research Institute, Sadashivanagar, Bangalore, India 560080, anish@rri.res.in}

\author{C. G. De Pree}
\affil{Department of Physics and Astronomy, Agnes Scott College, \\ 
141 E. College Ave., Decatur, GA 30030, USA}

\author{W. M. Goss}
\affil{National Radio Astronomy Observatory, P.O. Box O, Socorro, NM 87801, USA;
mgoss@nrao.edu}

\and

\author{K. R. Anantharamaiah\altaffilmark{1}}
\affil{Raman Research Institute, Sadashivanagar, Bangalore, India 560080}

\altaffiltext{1}{Deceased 29 October, 2001}

\begin{abstract}

We have detected C91$\alpha$ (8.5891 GHz) emission toward 4 ultra-compact \HII\ regions
(\UCHII s; W49G, J, L \& C) in the W49 North massive star forming region with 
the Very Large Array (VLA) at 3\arcsec\ resolution. No carbon line emission
was detected toward \UCHII s W49F, A, O, S and Q at this frequency to a
3$\sigma$ level of 2 mJy. We also observed the same region in the 
C75$\alpha$ line (15.3 GHz) with no detection at a 3$\sigma$ level of 6 mJy
with a 1\arcsec.7 beam. Detection of line emission toward these sources 
add supporting data to the earlier result of 
\nocite{retal05a}Roshi \etal\ (2005a) that many \UCHII s have
an associated photo-dissociation region (PDR). Similarity of the 
LSR velocities of carbon recombination lines and H$_2$CO absorption
toward \UCHII s in W49 North suggests that  
the PDRs reside in the dense interface zone surrounding these
\HII\ regions. Combining the observed carbon line parameters at 8.6 GHz
with the upper limits on line emission at 15.3 GHz, we obtain constraints on the
physical properties of the PDRs associated with W49G and J. The upper limit
on the number density of hydrogen molecule obtained from 
carbon line models is $\sim$ $5 \times 10^6$ \cmthree.
\end{abstract}


\section{Introduction}
\label{sec:intro}

Observations of molecular material in the vicinity of ultra-compact \HII\ regions
(\UCHII s) have shown that the ambient molecular gas may  
have very high densities. For example, observations of high density molecular 
tracers in the W49 North molecular cloud core show that 
the molecular density is a few times 10$^6$ \cmthree 
(\nocite{sgs93}Serabyn, G\"{u}sten \& Schulz 1993; 
see also \nocite{c02}Churchwell 2002 and references therein). 
The presence of such high density gas surrounding the \UCHII s has other 
effects. In particular, the far-ultra violet (FUV; 6 -- 13.6 eV) photons 
that escape the \UCHII\ produce photo-dissociation regions (PDRs) at
the interface between the ionized and molecular material (\nocite{ht97}
Hollenbach \& Tielens 1997). In these PDRs, carbon will be ionized since it
has a lower ionization potential (11.3 eV). The physical conditions in such
PDRs are ideal to produce observable carbon radio recombination lines (RRLs).  

A recent search for carbon RRLs toward \UCHII s have detected lines in
a majority of directions (\nocite{retal05a}Roshi \etal\ 2005a). These
observations were done near 9 GHz toward 17 \UCHII s and carbon RRLs were 
detected from 11 sources (65 \% detection rate). The large detection rate
indicates that most \UCHII s have an associated PDR. High-angular
resolution, multi-frequency carbon RL observation from these PDRs are
useful in addressing important issues related to \UCHII s and their evolution.
For example, such a data set can be used : (1) to estimate the 
physical properties of the PDR (\nocite{retal05a}Roshi \etal\ 2005a, b;
\nocite{getal98}Garay \etal 1998, \nocite{nwt94}Natta, Walmsley \& 
Teilens 1994) and (2) to address the `lifetime problem' of \UCHII s 
(\nocite{wc89}Wood \& Churchwell 1989) by investigating whether
some \UCHII s are pressure confined (\nocite{drg95}De Pree, Rodr\'iguez \& 
Goss 1995; \nocite{retal05a}Roshi \etal\ 2005a). 

In this paper, we report the detection of C91$\alpha$ (8.5891 GHz) 
emission toward several \UCHII s in W49 North. The W49 star forming region
lies close to the Galactic plane ($l$ = 43\ddeg.2, $b$ = $+$0\ddeg.00)
and is embedded in a giant molecular cloud of size $\sim$ 100 pc 
(\nocite{setal01}Simon \etal\ 2001), at a distance of 
11.4$\pm$1.2 kpc (\nocite{gmr92}Gwinn, Moran \& Reid 1992). Radio
continuum observations have identified 45 \UCHII s within the 
molecular cloud (\nocite{dmg97}De Pree, Mehringher Goss 1997). 
The properties of these \UCHII s were well studied 
through radio continuum as well
as hydrogen and helium RL observations (\nocite{dmg97}De Pree \etal\ 1997;
\nocite{detal00}De Pree \etal\ 2000). Here we
present new carbon RL observations of W49 North with the Very
Large Array (VLA).  The details of the observations 
are given in \S\ref{sec:obs}. In \S\ref{sec:res}, we provide the
parameters of the observed carbon lines and present models for
the line formation. A discussion of the results obtained is 
given in \S\ref{sec:dis}. 

\section{Observation and data reduction}
\label{sec:obs}

The observations were carried out on 22 November 1998 using the 
Very Large Array of the National Radio Astronomy Observatory\footnote
{The National Radio Astronomy Observatory is a facility of
the National Science Foundation operated under a cooperative
agreement by Associated Universities, Inc.}.
We used the C array for the observations. The helium and carbon RRLs 
at both 8.6 GHz ( He and C 91$\alpha$ ) and 15.3 GHz (He and C 75$\alpha$) 
were interleaved during the 8 hour period. The synthesized beam is 
1\arcsec.8 $\times$ 1\arcsec.6 (PA: $-$18\ddeg) at 15.3 GHz and  
3\arcsec.1 $\times$  2\arcsec.9 (PA: $-$17\ddeg)
at 8.6 GHz. The observations were centered at 25 \kms (LSR) to center the 
recombination lines between the He and C lines ( the velocity 
difference between He and C is about 30 \kms ). The observed
field was centered at the coordinates 
RA (J2000) = 19$^h$10$^m$13$^s$.41 and 
$\delta$ (J2000) = 09\ddeg01$^{'}$16${''}$.1. 
We used a total bandwidth of 3.125 MHz with 127 channels spaced by 24.4 kHz 
(velocity resolution $\sim$ 1 \kms) at 8.6 GHz and 6.25 MHz with 63 channels spaced 
by 97.7 kHz (velocity resolution $\sim$ 2 \kms) at 15.3 GHz. Two orthogonal circular 
polarizations were observed. The source 3C286 was used as the absolute 
flux density calibration and phase calibration was carried out by 
frequent observations of the compact source B1920+154. The bandpass 
response was determined by observations of 3C84 and 3C273.

The images were made using natural weighting in order to improve the 
sensitivity to extended sources. The continuum was subtracted using 
UVLSF in AIPS based on the line free channels. The final line cubes 
have an RMS noise of 0.7 mJy /beam (Hanning smoothed) at 8.6 GHz 
and 1.9 mJy /beam at 15.3 GHz.

\section{Results}
\label{sec:res}

\subsection{Carbon recombination line emission}

\begin{figure}
\plotone{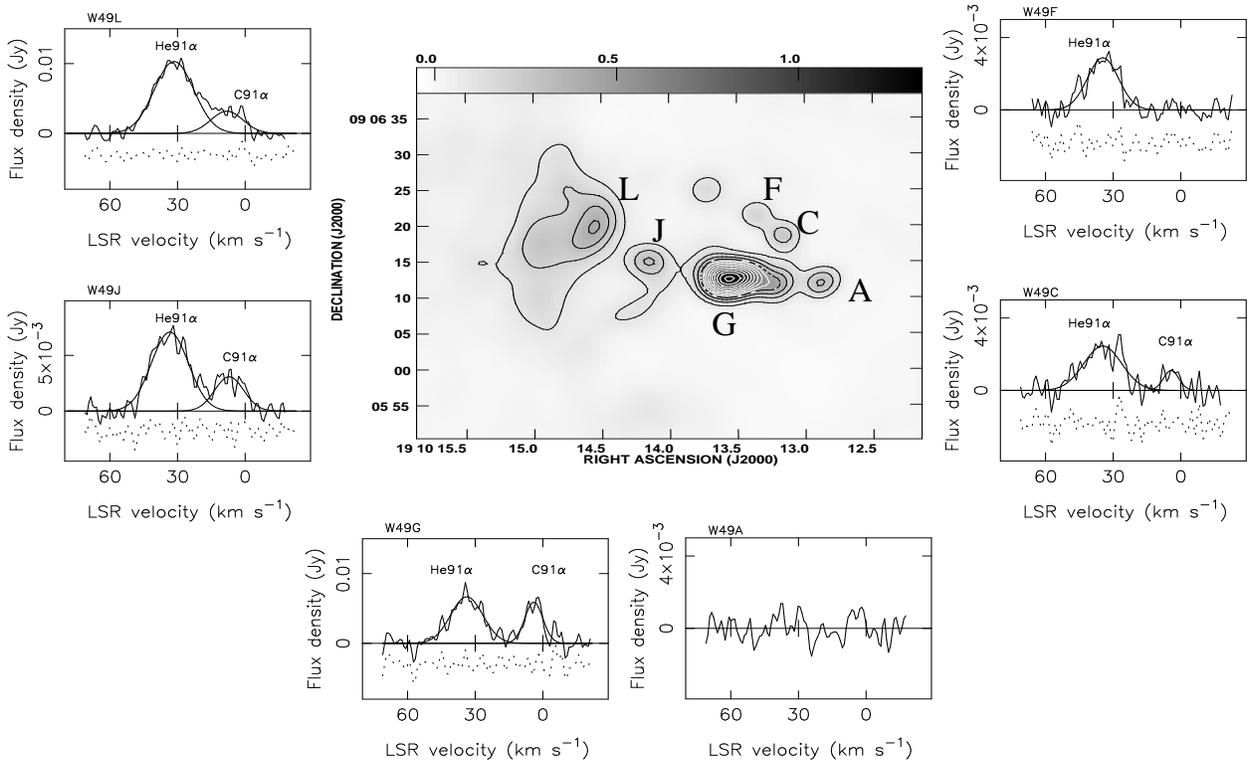}
\caption{
8.6 GHz continuum image of W49 North and spectra toward 6 \UCHII s. 
The angular resolution of the image is 3\arcsec.1 $\times$ 2\arcsec.9
(beam position angle $-$17\ddeg). The RMS noise in the continuum image 
is 7 mJy/beam. The contour levels in the image are 
(1, 2, 3, 4, 5, 6, 7, 8, 9, 10, 11, 12, 13) $\times$ 100 mJy/beam and
the grey scale ranges from $-$0.041 to 1.332 Jy/beam. We adopt the
naming convention followed by Dickel \& Goss (1990) to designate the
\UCHII s. The spectra are obtained by averaging the data over a region
specified in Table~\ref{tab1}.  
The fitted Gaussian components to the line emission
are also shown in each spectrum. Carbon line is detected toward
W49G and weak lines are present in the spectra toward W49J, C and L.
Spectra toward W49F and A are examples of our non-detections of carbon lines. 
\label{fig1} } 
\end{figure}

The central image in Fig~\ref{fig1} shows the 8.6 GHz continuum at 
$\sim$3\arcsec\ resolution. The RMS noise in the continuum image
is 7 mJy/beam. The spectra at 8.6 GHz obtained
toward six sources are also shown in figure~\ref{fig1}. The C91$\alpha$
line was detected toward four \UCHII s (W49G, J, C \& L). The
central coordinates and the extent of the region over which
the data are averaged are listed in
Table~\ref{tab1}.  Table~\ref{tab2} gives the
parameters of both carbon and helium lines obtained from the six spectra. 
The strongest carbon line (flux density 6 mJy) is observed toward W49G. 
This result is expected since the carbon line emission is dominated by 
stimulated emission at these frequencies (\nocite{nwt94}Natta \etal\ 1994; 
Roshi \etal\ 2005a, b) and W49G is the strongest continuum source
in the observed field. Weak carbon line emission is observed
toward W49J, C and L. The parameters of carbon line emission 
toward W49L may be affected by the larger velocity extent of helium 
line emission.  No carbon line emission is detected in our 15.3 GHz data
to a 3$\sigma$ level of 6 mJy with a beam size of 1\arcsec.7.

\subsection{Comparison with molecular line emission}

We can compare carbon RRL emission with existing CS (\nocite{sgs93}Serabyn \etal\ 1993)
and H$_2$CO (\nocite{dg90}Dickel \& Goss 1990) observations.
Multi-transitional observations of CS have provided
information on the structure of molecular cloud in W49 North 
and its kinematics. The cloud core consists primarily of three molecular clumps, 
designated as CS-NE, CS-C and CS-SW, with sizes 
between 0.4 -- 1 pc. Combining the CS observations with
H$_2$CO absorption studies, \nocite{sgs93}Serabyn \etal\ (1993) 
suggested that W49G, C and F are associated with the clump 
CS-C; W49A and L are probably associated with CS-SW and CS-NE, respectively. The
association of W49J with any one of three molecular clumps is not certain. 
Here we focus on the associations of \UCHII s in W49 North which are 
relevant for the present discussion. The LSR velocity of CS lines
from clump CS-C is 4.3$\pm$1.0 \kms, similar to the velocity of the
carbon RRLs (4.2$\pm$1.0 \kms) in the direction of W49G and C. 
These two sources also
exhibit H$_2$CO absorption components with an average velocity 
of 5.2$\pm$0.8 \kms. The carbon RRLs observed from W49L and J have  
velocities of 7.1$\pm$0.6 \kms, different from the velocity
of the CS line from the clump CS-NE (11.9$\pm$1.0 \kms). 
The H$_2$CO absorption components observed toward
these sources have velocities 7.2$\pm$0.2 \kms 
(\nocite{dg90}Dickel \& Goss 1990; We estimated approximate errors 
on the central velocity of H$_2$CO absorption line from the data
given in the reference). 
The similarity of the velocity of the H$_2$CO absorption 
components and carbon RRL emission toward all four sources
suggests that the PDR producing carbon lines reside in the gas 
responsible for H$_2$CO absorption.

\subsection{Models for carbon line emission}
\label{sec:cmodel}

We present models for carbon line emission toward W49G and J.
Constraints on the physical properties of the PDR  
could not be obtained toward W49C because the line emission
is weak. The W49L data are excluded from modeling
since the line parameters may be affected by the large velocity extent
of the helium line emission. 
The detailed steps involved in modeling the carbon line emission are described by 
\nocite{retal05b}Roshi \etal\ 2005b.  A homogeneous 
`slab' of PDR material placed in front of the \UCHII\ is
considered for modeling. The radiative transfer equation 
for non-LTE cases is solved to obtain the RL  
flux density. We used the computer code originally developed 
by \nocite{bs77}Brocklehurst \&  Salem (1977) and later modified 
by \nocite{ww82}Walmsley \& Watson (1982) and 
\nocite{pae94}Payne, Anantharamaiah \& Erickson (1994) to
calculate the non-LTE departure coefficients.

\begin{figure}
\plotone{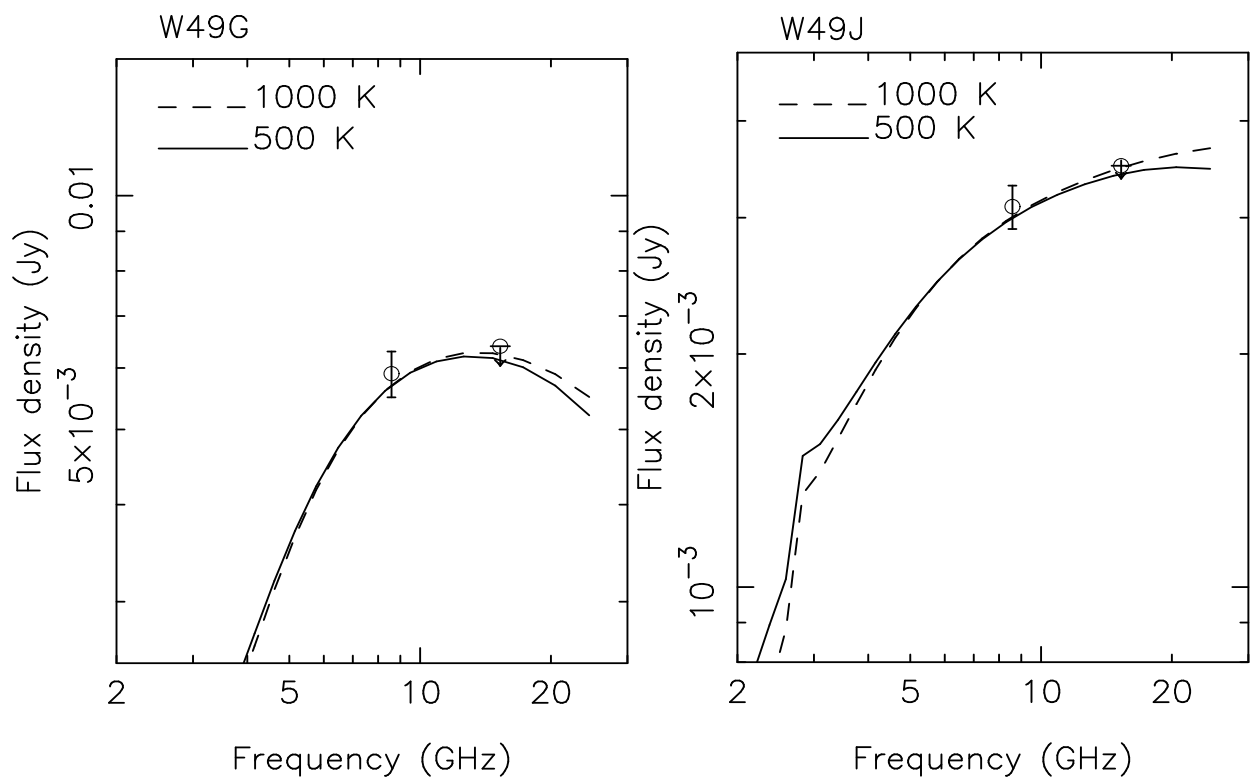}
\caption{
Carbon line flux density as a function of frequency from PDR 
models (see \S\ref{sec:cmodel}) toward W49G (left) and J (right).
The solid and dashed curves correspond to gas temperatures of 500 and 
1000 K respectively. The electron densities and path lengths for 
these models are given in Table~\ref{tab3}.  The upper limits 
on carbon RL emission at 15.3 GHz and C91$\alpha$ flux densities 
with $\pm$ 1$\sigma$ error are also marked. 
\label{fig2} } 
\end{figure}

The abundance of carbon required for the calculation
of departure coefficient is taken as 3 $\times$ 10$^{-4}$, 
which implies a depletion factor of 25 \% (\nocite{nwt94}Natta \etal 1994).
The departure coefficients also depend on the
background radiation field, assumed to be a thermal background 
from the \UCHII. The continuum emission at 8.6 and 15.3 GHz from the 
\UCHII\ (see Table~\ref{tab2}) is used to estimate an average emission measure 
for an assumed temperature of 9000 K. The estimated emission measures
for W49G and J are summarized in Table~\ref{tab3}. 
For a given PDR gas temperature, we varied the PDR emission measure 
to obtain the carbon line flux densities which are consistent with 
the 8.6 GHz detection and 15.3 GHz upper limit. 
Fig~\ref{fig2} shows the results of modeling for W49G and J.
The data at the two frequencies provide an upper limit on the PDR density 
and lower limit on the line of sight thickness of the PDR. These
limits are obtained for assumed temperatures of 500 and 1000 K 
(\nocite{nwt94}Natta \etal 1994, \nocite{retal05b}Roshi \etal\ 2005b) 
and summarized in Table~\ref{tab3}. The electron density obtained
from modeling is converted to molecular density $n_{H_2}$ 
using the assumed abundance of gas phase carbon and by
assuming that hydrogen is in molecular form in the PDR. 

\section{Discussion}
\label{sec:dis}

The detection of carbon RRLs toward a number of sources in 
the star forming region W49 North confirm the conclusion  
(\nocite{retal05a}Roshi \etal\ 2005a) that most \UCHII s may well have an 
associated dense PDR. These PDRs are formed at 
the interface between the ionized and dense neutral material,
perhaps a dusty zone separating the two region as in
the case of OH maser sources W49B and G (\nocite{dg90}Dickel 
\& Goss 1990). These interface zones are dense with $H_2$
densities $\sim$  10$^5$ \cmthree. Observational evidence
for the presence of such zones surrounding W49C and J 
exists; for W49L the existence of a distinct 
interface region is uncertain. 
\nocite{dg90}Dickel \& Goss (1990) suggest
that the interface zone is responsible for the H$_2$CO absorption. 
Carbon RRLs observed toward the \UCHII s in W49 North
have LSR velocities similar to that of  
H$_2$CO absorption components. This similarity suggests that the 
PDR responsible for carbon line emission resides in 
these interface zones surrounding
\UCHII s. Modeling the carbon line emission has provided 
upper limits on $H_2$ densities, which is 
$\sim$ 5 $\times$ 10$^6$ \cmthree. 
This upper limit is consistent with the densities inferred for
the molecular gas in the interface zone from H$_2$CO absorption studies
(\nocite{dg90}Dickel \& Goss 1990).




\begin{deluxetable}{cccc}
\tablecolumns{4}
\tablewidth{0pc}
\tablecaption{Details of the region over which the spectra are averaged 
\label{tab1}}
\tablehead{
Source & RA(J2000)\tablenotemark{a} & DEC(J2000)\tablenotemark{a} & Avg. area\tablenotemark{a}  \\
       &($^h$ $^m$ $^s$)            & ($^o$ $^{'}$ $^{''}$)       &  (arcsec$^2$)} 
\startdata
W49G & 19 10 13.54& 09 06 12.6 & 19  \\
W49J & 19 10 14.15& 09 06 15.2 & 14  \\
W49L & 19 10 14.44& 09 06 21.2 & 16  \\ 
W49C & 19 10 13.13& 09 06 18.6 & 11  \\ 
W49F & 19 10 13.34& 09 06 22.1 &  9  \\
W49A & 19 10 12.84& 09 06 12.1 & 12 \\
W49O & 19 10 16.35& 09 06 07.3 & 18 \\
W49S & 19 10 11.74& 09 05 27.0 & 27 \\
W49Q & 19 10 10.56& 09 05 13.9 & 50 \\
\enddata
\tablenotetext{a}{The spectrum for each source in Fig.~\ref{fig1}
and line parameters in Table.~\ref{tab2} are 
obtained by averaging the data over the area given here. The central
coordinates of the area are also given.}
\end{deluxetable}


\begin{deluxetable}{cccccccc}
\tabletypesize{\small}
\tablecolumns{7}
\tablewidth{0pc}
\tablecaption{Parameters of the observed recombination lines
\label{tab2}}
\tablehead{
\colhead{Source} &$f_{obs}$ & $S_c$\tablenotemark{a} & \colhead{Line} & \colhead{$S_L$\tablenotemark{b}} & \colhead{$\Delta V $\tablenotemark{b}} &\colhead{$V_{LSR}$\tablenotemark{b}} \\
   &\colhead{(GHz)} & \colhead{(Jy)}  &                & \colhead{(mJy)} & \colhead{(\kms)}      &\colhead{(\kms)}  } 
\startdata
W49G & 8.6 & 1.9(0.3) & C91$\alpha$  & 5.9 (0.4) & 9.5 (0.8) & 4.0(0.3)    \\
     &     &          & He91$\alpha$ & 6.7 (0.3) & 16.8(0.9) & 6.4(0.4)    \\
     & 15.3 & 2.4(0.3) & C75$\alpha$  & (6.4)     &           &             \\
W49J & 8.6 & 0.3(0.1)& C91$\alpha$  & 3.1 (0.2) & 15.1(1.5) & 7.1(0.6)    \\
     &  &   & He91$\alpha$ & 7.1 (0.2) & 19.3(0.8) & 6.0(0.3)  \\
     & 15.3 & 0.3(0.1)& C75$\alpha$  & (3.5)     &           &             \\
W49L & 8.6 & 0.5(0.1)& C91$\alpha$  & 3.2\tablenotemark{c} (0.2) & 16.7(1.9) &  7.1(0.6) \\
     &  &   & He91$\alpha$ & 10.3(0.2)  & 20.7(0.7)& 4.1(0.3)  \\
W49C & 8.6 & 0.2(0.1)& C91$\alpha$  & 1.1\tablenotemark{d} (0.2) & 8.0       & 4.3(0.9) \\
     &  &   & He91$\alpha$ & 2.4(0.2)  & 19.8(1.4) & 7.2(0.4)  \\
W49F & 8.6 & 0.13(0.03)& C91$\alpha$  &  (0.4)    &           &            \\
     &  &   & He91$\alpha$ & 2.9(0.1)  & 16.2(0.9) & 7.1(0.4)  \\
W49A & 8.6 & 0.2(0.1)&              &  (0.7)    &           &           \\
W49O & 8.6 & 0.5(0.1)& C91$\alpha$  &  (0.7)    &           &            \\
     &  &   & He91$\alpha$ & 4.2(0.2)  & 29.6(1.4) & 5.9(0.6)  \\
W49S & 8.6 & 0.5(0.1)& C91$\alpha$  &  (0.7)    &           &            \\
     &  &   & He91$\alpha$ & 2.2\tablenotemark{d}(0.2) & 26.4(2.6) & $-$3.0(1.1)  \\
W49Q & 8.6 & 0.5(0.1)& C91$\alpha$  &  (1.3)    &           &            \\
     &  &   & He91$\alpha$ & 7.5(0.2)  & 17.4(1.1) & 8.7(0.5)  \\
\enddata
\tablenotetext{a}{Continuum flux density with 1$\sigma$ error obtained over the area given in Table~\ref{tab1}.}
\tablenotetext{b}{The line parameters are obtained from the spectrum averaged over
the area given in Table~\ref{tab1}. The errors on the parameters are 1$\sigma$ values.}
\tablenotetext{c}{Carbon line parameters are affected by the large velocity extent 
of the helium line emission. }
\tablenotetext{d}{Tentative detection.}
\end{deluxetable}


\begin{deluxetable}{llcccc}
\tabletypesize{\small}
\tablecolumns{6}
\tablewidth{0pc}
\tablecaption{Results of modeling 
\label{tab3}}
\tablehead{
\colhead{Source} & $EM_{bg}$\tablenotemark{a} & \colhead{T$_{PDR}$} & \colhead{n$_e^{PDR}$} & \colhead{$l$} &\colhead{$n_{H_2}$}  \\
                 & (pc cm$^{-6}$) &\colhead{(K)}   & \colhead{(cm$^{-3}$)} & \colhead{($\times$ 10$^{-4}$ pc)} & \colhead{($\times$ 10$^{6}$ cm$^{-3}$)} } 
\startdata
W49G & 6.5$\times$10$^7$ & 500  & $<$ 3000 & $>$ 0.7  & 5.0 \\
     &                   & 1000 & $<$ 3500 & $>$ 1.9  & 6.0  \\
W49J & 1.2$\times$10$^7$ & 500  & $<$2800 & $>$ 3.6  &  4.5  \\
     &                   & 1000  & $<$3000 & $>$ 10.1  & 5.0   \\
\enddata
\tablenotetext{a}{Estimated emission measure of the \UCHII\ using
the continuum data for an assumed electron temperature of 9000 K.} 
\end{deluxetable}

\end{document}